\documentclass[12pt]{article}
\usepackage{url}
\usepackage{xcolor}
\usepackage{eepic}
\usepackage{graphicx}

\usepackage{bm}
\RequirePackage{times}
\RequirePackage{mathptm}
\RequirePackage{courier}

\usepackage{longtable}
\usepackage{dcolumn}

\begin{document}


\title{Is Feasibility in Physics Limited by Fantasy Alone?\thanks{To appear in H.~Zenil (ed.). {\em A Computable Universe: Understanding Computation \& Exploring Nature as Computation},
 World Scientific,
 2012.}}



\author{Cristian S.  Calude\thanks{Computer Science Department,
The University of Auckland,
 Private Bag 92109,
 Auckland,
 New Zealand.},  \,
 Karl Svozil\thanks{Institute for Theoretical Physics, Vienna University of Technology,
Wiedner Hauptstra\ss e 8-10/136, A-1040 Vienna, Austria.}}

\maketitle

\begin{abstract}
Although various limits on the predicability of physical phenomena as well as on physical knowables are commonly established and accepted, we challenge their ultimate validity. More precisely, we claim that fundamental limits arise only from our limited imagination and fantasy. To illustrate this thesis we
give evidence that the well-known Turing incomputability barrier can be trespassed via quantum indeterminacy.
From this algorithmic viewpoint, the  ``fine tuning'' of physical phenomena amounts to a ``(re)programming'' of the universe.\end{abstract}


Take a few moments for some anecdotal recollections.
Nuclear science has made true the ancient alchemic dream of producing
gold from other elements such as mercury through nuclear reactions.
A century ago, similar claims would have disqualified anybody presenting them as quack.
Medical chemistry discovered antibiotics which cure Bubonic plague, tuberculosis, syphilis, bacterial pneumonia,
as well as a wide range of bacterial infectious diseases
which were considered untreatable only one hundred years ago.
For  contemporaries it is hard to imagine the kind of isolation, scarcity in international communication,
entertainment and transportation most of our ancestors had to cope with.

This historic anecdotal evidence suggests that what is considered tractable,
operational and feasible depends on time.
One could even extend speculations to the point where everything that is imaginable is also feasible.
In what follows, we shall concentrate on some physical issues which might turn out to become relevant in the
no--so--distant future, and which might affect the life of
the generations succeeding ours to a considerable degree.
In particular, we shall consider the connections between time, space and the limit velocity of light in vacuum;
we shall ponder upon measurement;
and we shall discuss physical indeterminism and randomness, and  its relations to the possibility of trespassing the Turing incomputability barrier.

\section{Space-time}

One of the findings of special relativity theory is the impossibility to trespass the speed of light barrier ``from below;''
i.e., by starting out with subluminal speed.
This fundamental limit applies also to communication and information transfer.
Amazingly, this holds true even when quantum mechanics and ``nonlocal quantum correlations'' are taken into account,
stimulating a notion of ``peaceful coexistence'' between quantum mechanics and special relativity theory.
Thereby, superluminal particles,
as well as the inclusion of field theoretic effects such as an index of refraction smaller than unity,
supercavitation in the quantum ether,
or general relativistic effects by locally rotating masses,
wormholes or local contraction and expansion of space-time,
possibly also related to time travel,
to name but a few, cannot be  excluded {\it a priory.}



Recent operational definitions of space-time and velocity, in order to physically represent the former, conventionalised the latter:
Initially,  the constancy of the velocity of light in vacuum in all reference frames was treated
as an empirical fact.  Since 1983 it has been
frame-invariantly standardised
by {\it Resolution Number 1} of the {\it  17th Conf{\'e}rence G{\'e}n{\'e}rale des Poids et Mesures (CGPM)}
in which the following SI {\it (International System of Units)} operational definition of the meter
has been adopted:
{\em ``The metre is the length of the path travelled by light
in vacuum during a time interval of 1/299 792 458 of a second.''}
As a result, the empirical fact associated with this convention,
as predicted by relativity theory,
is the proposition that the length of a solid body depends neither on its spatial orientation,
nor on the inertial frame relative to which that body is at rest~\cite{peres-84}.

Indeed, by a theorem of
incidence geometry~\cite{lester},
linear Lorenz-type transformations follow from the frame-invariant standardisation of the velocity of light alone
and appear to be a formal consequence of the conventions adopted by the SI.
In such an approach, the physics resides in the invariance of Maxwell's equations and the equations of motion in general,
as well as in the invariance
of all physical measures based on matter stabilised by them, such as the length or duration of a space or time scale.
This, after all, suits the spirit of Einstein's original 1905 paper,
which starts out with conventions defining simultaneity and then proceeds with kinematics and by unifying electric and magnetic phenomena.
Of course, for the sake of principle, everybody is free to choose other ``limiting speeds,''
thereby implicitly sacrificing the form invariant representation of the equations of motion in inertial frames dominated and stabilised by electromagnetic interactions.
In this way it would also not be difficult to adopt special relativity
to findings of higher signalling and travel speeds than the velocity of light in vacuum.


Since antiquity, natural philosophers and scientists have pondered about
the (in)finite divisibility of space-time, about its (dis)continuity, and
about the (im)possibility of motion.
In more recent times, the ancient Eleatic arguments ascribed to Parmenides and Zeno of Elea
have been revived to ``construct'' accelerated computations~\cite{weyl:49}
which serve as one of the main paradigms of the fast growing field of hypercomputation.

The most famous argument ascribed to Zeno is the impossibility for ``Achilles'' to overtake a turtle if the turtle is granted
to start some finite distance ahead of Achilles, even though the turtle moves, say, one hundred times slower than Achilles:
for in the finite time it takes Achilles to reach the turtle's start position, it has already moved away from it and is still a
(tenth of the original) finite distance apart from Achilles. Now, if Achilles tries to reach that new point in space, the turtle
has made its way to another point and is still apart from Achilles.
Achilles' vain attempts to reach and overtake the turtle could be considered {\it ad infinitum}; with him coming ever closer to the
turtle but never reaching it.
By a similar argument, there could not be any motion, because in order to move from one spatial point to another,
one would have first to cross half-distance; and in order to be able to do this, the half-distance of the
half-distance, $\ldots$ again {\it ad infinitum.}
It might seem that because of the infinite divisibility of space, unrestricted motion within it is illusory because of this impossibility.

The modern-day ``solution'' of this seemingly impossible endeavour to move ahead of a slower object
resides in the fact that it takes Achilles an ever decreasing outer (extrinsic, exterior) time to reach
the turtle's previous position; so that if one takes ``the limit''
by summing up all infinitely many outer space and time intervals,
Achilles meets (and overtakes) the turtle in finite outer time,
thereby approaching an infinity of space-time points.
Of course, Achilles' approach is even then modelled by an infinite number of steps or trip segments,
which can be used to create an inner (intrinsic) discrete temporal counter.
If this inner counter could in some form be associated with the cycle of an otherwise conventional universal computer such as a universal
Turing machine or a universal cellular automaton~\cite{zuse-67,wolfram-2002},
then these ``machines'' might provide ``oracles'' for ``infinite computations.''
In this respect, the physics of space and time, and computer science intertwine.

The  accelerated Turing machine (sometimes called Zeno machine) is a Turing machine working in a
 computational space analogue to Zeno's scenario.  More precisely, an
accelerated Turing machine  is a Turing
machine that operates in a universe with two clocks:
for  the exterior clock each step is executed in a unit of time (we assume that steps are in some sense identical except for the time
taken for their execution) while for
the inner clock it takes an ever decreasing amount of time, say $2^{-n}$ seconds,  to perform its $n$th step.
Accelerated Turing machines have been implicitly described by Blake~\cite[p.~651]{Blake26} as well as Weyl~\cite[pp.~41-42]{weyl:49},
and studied in many papers and books since then.

Because an accelerated Turing machine can run an infinite number of steps
 (as measured by the exterior clock)
in one unit of time (according to the inner clock),
such a mechanism  may compute incomputable functions, for example, the characteristic function of
the halting problem.

How feasible are these
types of computation? This is not an easy question,
so not surprisingly  there is no definitive answer.
One way to look at this question is to study the relation between
 computational time and space. As expected, there is a similarity between computational time and space; however, this
parallel is not perfect. For example, it is not true that an accelerated
Turing machine which uses unbounded space has to use an infinite space for
some input. An accelerated Turing machine that uses a finite space (not necessarily bounded)
for all inputs  computes a  {\it computable function} (the function is not
   necessarily computed by the same machine)~\cite{calude-staiger-09}. Hence, if an accelerated Turing machine
   computes an incomputable function, then the machine has to use an infinite set of configurations  for infinitely many inputs.
Re-phrasing, going beyond Turing barrier with an accelerated Turing machine requires an infinite computational space
(even if the computational time is finite);
the computational space can be bounded (embedded in the unit interval),
 but cannot be made finite. Do we have such a space?
Maybe relativistic computation offers a physical model for hypercomputation~\cite{andreka-2009}.

\section{Measurement}


Another challenging question has emerged in the quantum mechanical context
but it equally applies for all reversible systems:
what is an irreversible measurement?
Because if  the quantum evolution is uniformly unitary and thus strictly reversible,
what is to be considered the separated ``measurement object'' and the ``measurement apparatus''
can be ``wrapped  together'' in a bigger system containing both, together with the ``Cartesian cut;'' i.e., the environment supporting communication between these two entities.
Any such bigger system is then uniformly describable by quantum mechanics, resulting in total reversibility of whatever
might be considered intrinsically and subjectively as a ``measurement.''
This in turn results in the principal impossibility of any irreversible measurement
(not ruling out  decoherence ``fapp;'' i.e., for all practical purposes);
associated with the possibility to ``reconstruct'' a physical state prior to measurement;
and to ``undo'' the measurement~\cite{hkwz}.
The quantum state behaves just as in Schr\"odinger's interpretation of the $\Psi$ function  as
a {\em catalogue of expectation values:} this catalogue can only be ``opened and read'' at a single page;
yet it may be ``closed'' again by ``using up'' all knowledge obtained so far, and then reopened at another page.

 Two related types of unknowables which have  emerged in the quantum context are
complementarity and value indefiniteness.
{\em Complementarity} is the impossibility to measure two or more observables instantaneously with arbitrary accuracy:
in the extreme case, measurement of one observable annihilates the possibility to measure another observable, and {\it vice versa.}
Despite attempts to reduce this feature to a ``completable'' incompleteness by Einstein and others,
and thus to preliminary, epistemological deficiencies of the quantum formalism,
the hypothetical ``quantum veil,''
possibly hiding the ``physical existence'' of the multitude of all conceivable (complementary) observables,
has maintained its impermeability until today.

As new evidence emerged,  the lack of classical comprehensibility  has gotten even worse:
whereas quasi-classical systems---such as generalised urn or finite automaton models~\cite{svozil-2008-ql}---feature complementarity, some
quantised systems with more than two measurement outcomes cannot be thought of as possessing any global ``truth function.''
As the Kochen-Specker theorem~\cite{specker-60,kochen1} shows,
they are {\em value indefinite} in the sense that there exist  (even finite) sets of observables which,
under the hypothesis of non-contextuality,
 cannot all (for some this might still be possible) have definite values
independent of the type of measurement
actually being performed.

Faced with the formal results, some researchers prefer a resolution in terms of {\em  contextual} realism:
measurement values ``exist'' irrespective of their ``actual measurement,''
but they depend on what other observables are measured alongside of them.
Another possibility is to abandon classical omniscience
and assume that an ``elementary'' quantum system is only capable
of expressing a {\em single} bit (or dit for $d$ potential measurement outcomes)~\cite{zeil-99}
or context; all other conceivable measurements are mediated by a measurement apparatus capable of
context translation.

\section{Indeterminacy and hypercomputability}

In the Pythagorean tradition, the universe computes.
Thus any method and measure to change its behaviour amounts to (re)programming.
If one remains within this metaphor, the character and ``plasticity'' of the ``substratum'' software and hardware needs to be exploited.
Presently, the Church-Turing thesis confines the universe to universal computability formalised by
recursion theory,
but is it conceivable that some physical processes transcend this realm?

Arguably, the most (in)famous  result in theoretical computer science is  Turing's
theorem saying that it is undecidable to determine whether a
general
computer program will halt or not.   This is formally known as the \emph{%
halting problem}.  More precisely, there is no computer program {\tt halt}
which given as input an arbitrary program {\tt p} runs a finite-time computation and returns 1 if
{\tt p} eventually stops and 0 if {\tt p} never stops (here we use a fixed universal Turing machine
to run programs).

There are two essential conditions imposed on {\tt halt}: a) {\tt halt} has to stop on every input, b)
{\tt halt} returns the correct answer.
It is easy to construct a program {\tt halt} that satisfies the above two
conditions for many very, very large sets of programs, even for infinite sets of programs,
but {\it not}, as Turing proved, for {\it all} programs.

So, one way to trespass the Turing barrier is to provide a
a physical mechanism which computes the function  {\tt halt} discussed above. There are many
proposals for such devices..
Let's first present a negative result:  using an information-theoretic
argument,  the possibility of having access to a time-travel machine
 would not solve the halting problem,
unless one
could travel back and forth in time at a pace exceeding the growth of any
computable function.

Would some quantum processes transcend the Turing barrier?
Surprisingly, the answer is yes~\cite{2008-cal-svo}, and the main reason is the incomputability
of quantum randomness.

In 1926, Max Born stated that (cf. \cite[p.~866]{born-26-1}, English translation in \cite[p.~54]{wheeler-Zurek:83})

\begin{quote}
{\em  ``From the standpoint of our quantum mechanics, there is no quantity
which in any individual case causally fixes the consequence of the collision;
but also experimentally we have so far no reason to believe that there are some inner properties of the atom
which condition a definite outcome for the collision.
Ought we to hope later to discover such properties [[$\ldots$]]  and determine them in individual cases?
Or ought we to  believe that the agreement of theory and experiment --- as to the impossibility
of prescribing conditions? I myself am inclined  to give up determinism in the world of atoms.''
}
\end{quote}

Born's departure from  the {\em principle of sufficient reason}
--- stating that every phenomenon has its explanation and cause ---
by postulating irreducible randomness~\cite{zeil-05_nature_ofQuantum} in the physical sciences did not specify formally the
type of ``indeterminism'' involved.
More recent findings related to the Boole-Bell, Greenberger-Horne-Zeilinger
as well as Kochen-Specker theorems for Hilbert spaces of dimension three onwards derive physical indeterminism
from value indefiniteness of at least one observable among finite complementary collections of observables.
As a result, although not necessarily all noncontextual observables in Kochen-Specker-constructions need to be value indefinite,
but at least one has to be.

Suppose that quantum value indefiniteness occurs uniformly and symmetrically distributed over all observables.
Because indeterminism and randomness are defined by algorithmic ``lawlessness'' and ``incompressibility''~\cite{chaitin:01}
any physical system featuring indeterminism and randomness cannot be simulated by a universal computer;
it ``outperforms'' any known computing machinery in terms of unpredictability.
With these assumptions, physical value indefiniteness and randomness  can thus be seen as valuable resources capable of serving as ``oracles''
for example, for Monte Carlo methods and primality testing requiring them.
Indeterminism becomes  an {\em asset} rather than a deficiency.

Contemporary realisations of quantum random number generators
involve beam splitters.
Thereby it should be noted that lossless beam splitters are reversible devices formalised by unitary transformations,
and that the single photons used constitute a two-dimensional Hilbert space which may be ``protected''
from cryptanalytic attacks ``lifting the hypothetic quantum veil'' by quantum complementarity only.
Indeed, it may not be totally unreasonable to point out that one of the greatest and mind-boggling quantum riddles of our time
is the rather ambivalent use of beam splitters:
on the one hand,   beam splitters are associated with random coin tosses, which ase
postulated to
yield absolute and irreducible randomness~\cite{zeil-05_nature_ofQuantum};
while on the other hand  beam splitters
are represented by dicrete reversible unitary operators;
the action of which could be totally reversed by serially composing two of them into a Mach-Zehnder interferometer.

Imperfections in measurements are typically corrected with von Neumann's procedure of normalisation --- ``compressing'' a bit sequence {\it via} the map
$00$,$11 \mapsto \{\}$ (00 and 11 are discarded),
$01 \mapsto 0$, and
$10 \mapsto 1$. The algorithm works under the hypotheses of
 independence and stationarity of the original sequence,
conditions which may not be satisfied in beam splitting experiments---for instance due to multiparticle statistics like the Hanbury Brown and Twiss effect.

Some quantum systems are protected by value indefiniteness grounded in the Kochen-Specker theorem
from Hilbert space dimensions three onwards.
As
the Kochen-Specker theorem requires complementarity, but the converse implication is not true, it follows that
 a system of two entangled photons in a singlet state
or systems with three or more
measurement outcomes may be more suitable for generating quantum random bits.
Obtaining more than two outcomes is not problematic as, if in a sequence of random elements drawn from an alphabet with $n>2$ symbols
 a fixed symbol is systematically removed, the resulting sequence is still random (over an alphabet of $n-1$
 symbols).

\section*{Final remarks}

There are exciting times ahead of us.
The limits which seem to be imposed upon us by various constraints
might decay into ``thin air'' as the conditions upon which these constraints are founded will lose their applicability and necessity, or even lose their operational validity.
Thus, we perceive physical tractability and feasibility wide open, positive, and full of unexpected opportunities.
Indeed, we just quiver at the extension of our imaginable ignorance; let alone the possibilities which we even lack to fantasise.
Any further scientific exploration of this realm has to be strongly encouraged.

\section*{Acknowledgement}
We thank A. Abbott and E. Calude for useful comments and criticism.


\end{document}